# Mechanochemical Synthesis and Magnetic Properties of the Mixed-Valent Binary Silver(I,II) Fluorides, $Ag^I_2Ag^{II}F_4$ and $Ag^IAg^{II}F_3$

Matic Belak Vivod[a,b], Zvonko Jagličić[c,d], Graham King[e], Thomas C. Hansen[f], Matic Lozinšek[a,b], and Mirela Dragomir[a,b]*

[a] *Jožef Stefan Institute, Jamova cesta 39, 1000 Ljubljana, Slovenia*
[b] *Jožef Stefan International Postgraduate School, Jamova cesta 39, 1000 Ljubljana, Slovenia*
[c] *Institute of Mathematics, Physics and Mechanics, 1000 Ljubljana, Slovenia*
[d] *Faculty of Civil and Geodetic Engineering, University of Ljubljana, Jamova cesta 2, 1000 Ljubljana, Slovenia*
[e] *Canadian Light Source, 44 Innovation Blvd, Saskatoon, SK S7N 2V3, Canada*
[f] *Institut Laue-Langevin, 38042 Grenoble Cedex 9, France*

*Corresponding author: mirela.dragomir@ijs.si

*Dedicated to the memory of Professor Boris Žemva, whose passion for fluorine chemistry ignited our pursuit in this field.*

**ABSTRACT**

Fluoridoargentates(II) represent a fascinating class of silver(II) compounds that exhibit structural and magnetic similarities to cuprate superconductors. However, their synthesis is challenging, leaving their properties largely unexplored. In this study, mechanochemistry is introduced as a new technique for the synthesis of fluoridoargentates(II), which avoids the use of anhydrous HF or heating, and employs simple equipment. Furthermore, by ball milling of commercially available precursors, AgF and $AgF_2$, we obtained the first two examples of binary mixed-valent silver(I,II) phases, $Ag^I_2Ag^{II}F_4$ and $Ag^IAg^{II}F_3$. While the $Ag^I_2Ag^{II}F_4$ phase was obtained at room temperature, the $Ag^IAg^{II}F_3$ phase is metastable and required milling under cryogenic conditions. Using synchrotron powder X-ray diffraction, it was found that $Ag^I_2Ag^{II}F_4$ crystallizes in the $P2_1/c$ space group and is isostructural to β-$K_2AgF_4$. Here, double-bridged $[AgF_6]^{4-}$ octahedra form chains that propagate along the *a*-crystallographic direction, giving a quasi-1D canted antiferromagnetic character, as shown by magnetic susceptibility. The $Ag^IAg^{II}F$ phase adopts the $P-1$ space group, is isostructural to $AgCuF_3$ and shows features of a one-dimensional antiferromagnet. It is expected that this facile synthetic approach will enable the expansion of silver(II) chemistry and accelerate the search for a silver analogue to cuprate superconductors.

**INTRODUCTION**

Systems containing $S = ½$ cations offer a rich playground to explore quantum effects such as spin liquids, spin-Peierls instabilities, superconductivity and the like.[1,2] Compared to other metal cations like $Cu^{2+}$, $Ti^{3+}$, or $V^{4+}$, systems containing $Ag^{2+}$ ($4d^9$) are less explored although they show great promise for very intriguing physics such as exotic magnetism and even superconductivity as cuprate analogues.[3,4] However, unlike $Cu^{2+}$, which is in a common oxidation state, $Ag^{2+}$ is very unstable and reactive species.[5,6] In oxides, $Ag^{2+}$ disproportionate to $Ag^+$ and $Ag^{3+}$. Fluorine is the only element sufficiently electronegative to readily stabilize Ag in a +2 oxidation state. These compounds, called fluoridoargentates(II), are truly exceptional in various aspects.[7] Even the most fundamental binary compound, $AgF_2$, displays intriguing structural and electronic properties similar to the parent compound of the cuprate superconductors, $La_2CuO_4$,[8] which features strong mixing of $Ag(4d^9)$ and $F(2p^2)$ orbitals, a remarkable covalence of the Ag−F bond, and significant magnetic superexchange in two dimensions.[9,10] Similar to $Cu^{2+}$, the $4d^9$ $Ag^{2+}$, $S = ½$, can also be found in a Jahn-Teller elongated or compressed octahedral environment. By analogy with cuprates, where superconductivity can be achieved by charge doping, mixed-valence silver fluorides with silver(I,II) or silver(II,III) fluorides could also be promising candidates to achieve superconductivity.[3] However, synthesizing these compounds is challenging and has hindered their exploration.

Currently, the binary fluorine–silver system includes six phases: $Ag_2F$,[11,12,13,14] AgF,[15,16] $AgF_2$,[17,18,19,20] $AgF_3$,[21,22] $Ag_2F_5$,[23,24] and $Ag_3F_8$.[25] In these compounds, silver adopts the +½, +1, +2 and +3 oxidation state. Interestingly, only two binary mixed-valence silver fluorides are known up to date: $Ag_2F_5$ or $Ag^{II}F[Ag^{III}F_4]$, and $Ag_3F_8$ or $Ag^{II}[Ag^{III}F_4]_2$. In both cases, silver is in a mixed +2/+3 oxidation state. Theoretical calculations, motivated by the promising properties of the silver(I,II) phases, were performed to search for the most stable silver fluoride polymorphs.[26,27,28] Two papers suggest two novel stoichiometries with $Ag^+$ and $Ag^{2+}$ cations, namely $Ag_3F_4$ and $Ag_2F_3$, which could be thermodynamically stable at ambient conditions.[26,27] The hypothetical $Ag^I_2Ag^{II}F_4$ or $Ag_3F_4$ phase was predicted to adopt three different space groups: a post-perovskite monoclinic $P2_1/c$ (β-$K_2AgF_4$ or $Na_2CuF_4$-type),[29] the orthorhombic $Cmca$ (similar to α-$K_2AgF_4$ and $La_2CuO_4$)[30] and the tetragonal $I4/mmm$ space group ($Nd_2CuO_4$-type).[31] With a combination of chemical intuition and DFT calculations, the first report considered the most stable polymorph with the lowest energy to be the monoclinic $P2_1/c$, whereas the orthorhombic hypothetical phase was found to be much higher in energy.[26] A follow-up theoretical investigation proposed only one polymorph with the $I\bar{4}2d$ space group ($CdMn_2O_4$-type) as the ground state structure of $Ag^I_2Ag^{II}F_4$ stable up to 19 GPa.[27]

The existence of another Ag(I,II) phase with the formula $Ag^IAg^{II}F_3$ or $Ag_2F_3$ was also predicted by DFT. In the first report, four different crystal structure models were evaluated for their stability at ambient conditions: $Pbnm$ ($GaFeO_3$-type structure), $Pmcn$ ($CuTeO_3$-type structure), $P\bar{1}$ ($NaCuF_3$-type structure) and $C222_1$ ($BaVS_3$-type structure).[26] The first three candidates were found to be equivalent in terms of energy of formation. Another theoretical study proposed only a polymorph with the $P2_1/m$ space group ($CaIrO_3$-type structure).[27]

The first theoretical report showed that the calculated energies of formation, $\Delta E_f$, for the most stable polytypes was in between −0.03 eV and −0.07 eV per formula unit for $Ag^IAg^{II}F_3$ and in between −0.17 and −0.21 eV for $Ag^I_2Ag^{II}F_4$.[26] The more negative energy of formation for $Ag^I_2Ag^{II}F_4$ compared to $Ag^IAg^{II}F_3$ indicates that $Ag^I_2Ag^{II}F_4$ is the thermodynamically stable product of a reaction between $AgF_2$ and AgF. Following these predictions, experimental efforts to synthesize these phases have fallen short, as current synthetic methods have not been able to provide access to these compounds which could be metastable. Common problems include the poor solubility of the main precursor $AgF_2$ in anhydrous HF and thermal decomposition of $AgF_2$ at temperatures < 400 °C.[4,32] Mechanochemistry could overcome the disadvantages of the conventional synthesis methods offering room-temperature reaction conditions, faster kinetics, and solvent-free synthesis.[33,34,35,36] Mechanochemistry represents a paradigm shift in the synthesis of materials that can enable the formation of metastable phases that are difficult to obtain using high-temperature methods.[37,38] While there are numerous reports on the high-energy milling of oxides[39] indicating that it is an established technique, the mechanochemistry of fluorides on the other hand is still in its early stages of development, with the mechanochemistry of fluoridoargentates(II) being virtually an unexplored area of research at the onset of this work. This study proves that mechanochemistry is an effective approach to the formation of the anticipated mixed-valent binary silver(I,II) fluorides and showcases that it could be implemented in the field of fluoridoargentates(II).

**RESULTS AND DISSCUSSION**

**$Ag^I_2Ag^{II}F_4$**

For the synthesis of $Ag^I_2Ag^{II}F_4$ a 2:1 molar ratio of AgF to $AgF_2$ was balled milled at room temperature. Further experimental details are provided in the **Supporting Information**, **SI**. The laboratory PXRD data (**Figure 1a)** indicated a possible successful reaction but also the presence of about 10 wt.% of AgF as a result of a partial decomposition of $AgF_2$ during milling. For comparison, a solid-state synthesis similar to that used for the synthesis of $M_2AgF_4$ reported in the literature[40] was also attempted. However, in the present study, a lower temperature of 300 °C was used to minimize the decomposition of $AgF_2$ at the expense of longer dwell times of up to three days. A comparison of the diffractograms between the ball milled and solid-state synthesized samples (**SI**, **Figure S1**) indicate that the samples prepared by ball milling show a substantial peak broadening compared to the solid-state synthesized sample, which indicates the presence of nm-sized crystallites and the existence of defects or microstrain induced during milling. A higher background is also observed, indicative of the presence of an amorphous phase. Similar observations were also noted in the milling of oxides.[39] For the structural refinement, high-resolution synchrotron data was collected on the pale light-brown polycrystalline sample prepared with the solid-state method due to the higher crystallinity of this sample (**Figure 1b**). Unassigned peaks that did not belong to unreacted reagents (AgF or $AgF_2$) were indexed to a monoclinic $P2_1/c$ unit cell (**SI**, **Figure S2**) with similar unit cell parameters to $Na_2CuF_4$.[29] This structure was then used as a starting model for a three-phase Rietveld refinement. It was expected that the resulted weight ratio of unreacted reagents (AgF and $AgF_2$) would be 1.74:1, but this was not the case, since the refinement indicated about 23 wt.% of AgF and only around 1 wt.% of $AgF_2$. It was assumed that the larger amount of AgF was a result of sample decomposition in the capillary after sealing for PXRD analysis, indicated by the loss of capillary transparency. Indeed, when the PXRD measurement was performed on a fresher sample, it revealed an AgF content of 14.3(1) wt.% (**SI**, **Figure S3**). This showcases the instability of the sample.

An experimental suggestion of the $Ag^I_2Ag^{II}F_4$ phase came also from Raman measurements performed on $AgF_2$ with a green excitation laser ($\lambda$ = 532 nm). An increase of the illumination time led to changes in the Raman spectrum such as a decrease in the 250 cm$^{-1}$ band and a significant increase in the 420 cm$^{-1}$ band (**SI**, **Figure S4**), indicating a possible photochemical decomposition of $AgF_2$. Similar observations were already reported without a confirmation of the formation of the $Ag^I_2Ag^{II}F_4$ phase.[41] This intense 420 cm$^{-1}$ band could be associated with the $[AgF_4]^{2-}$ units present in β-$K_2AgF_4$ and $Na_2AgF_4$, indicating structural similarities between the new phase and these compounds. Assuming that the new $Ag^I_2Ag^{II}F_4$ phase is isostructural to $Na_2AgF_4$ and β-$K_2AgF_4$, it would crystallize in the $P2_1/c$ space group and would have eighteen Raman active modes ($9A_g + 9B_g$) as predicted by the group theory.[42] Raman spectroscopy measurements performed on the 2:1 milled sample (**Figure 1c**) indicated only six peaks resembling those observed in the $Na_2AgF_4$ and β-$K_2AgF_4$ compounds. To make a direct comparison, β-$K_2AgF_4$ was also synthesized by ball milling. The PXRD analysis confirmed the successful formation of the β-$K_2AgF_4$ phase (**Figure S5**).

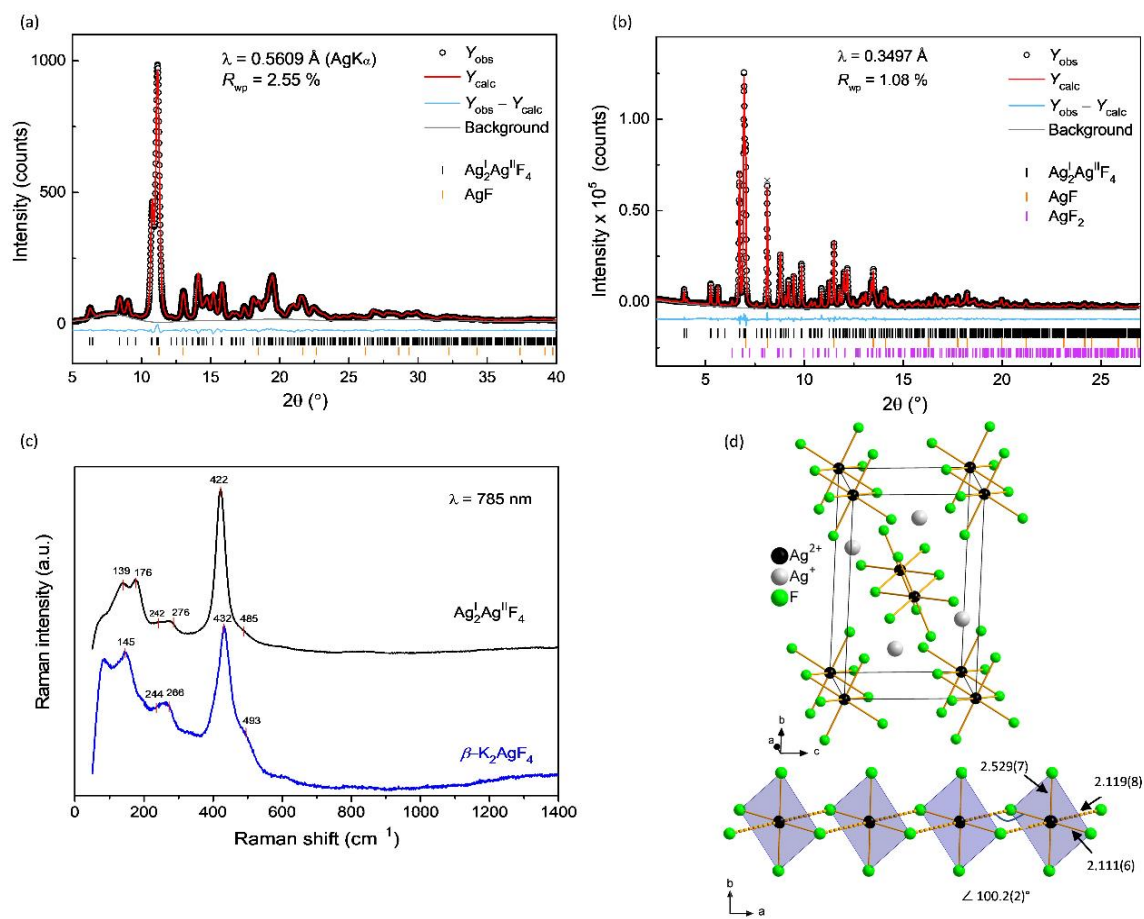

**Figure 1.** **(a)** Rietveld refinement of the laboratory PXRD data of mechanochemically synthesized $Ag^I_2Ag^{II}F_4$ collected at room temperature; wt.% ($Ag^I_2Ag^{II}F_4$) = 88.7(2); wt.% (AgF) = 11.3(6). **(b)** Three-phase Rietveld refinement of the synchrotron data collected on the $Ag^I_2Ag^{II}F_4$ sample prepared by solid-state method; wt.% ($Ag^I_2Ag^{II}F_4$) = 76.2(1); wt.% (AgF) = 22.9(1); wt.% ($AgF_2$) = 0.9(5). **(c)** Raman spectra of $Ag^I_2Ag^{II}F_4$ and β-$K_2AgF_4$ synthesized by ball milling. **(d)** The crystal structures of $Ag^I_2Ag^{II}F_4$ with the main structural motif being a chain-like composed of double-fluorine bridged $[AgF_6]^{4-}$ units.

Crystallographic parameters of $Ag^I_2Ag^{II}F_4$ obtained with the Rietveld refinement of synchrotron data are found in the SI (**Table S1**). The Bond Valence Sum (BVS) analysis was also performed (**Table S2**) confirming the mixed +1/+2 valence state of silver in this binary silver fluoride phase.

The Raman spectrum of β-$K_2AgF_4$ shows the most intense peak at 432 cm$^{-1}$ with a shoulder at 493 cm$^{-1}$ (**Figure 1c**), in good accordance with the literature reports.[43] The 432 cm$^{-1}$ Raman band can be attributed to the symmetric vibrations of the $[AgF_4]^{2-}$ subunits.[44] With an intense peak at 422 cm$^{-1}$ and a shoulder at 485 cm$^{-1}$, the Raman spectrum of the $Ag^I_2Ag^{II}F_4$ phase seems to be very similar to the β-$K_2AgF_4$ phase, confirming the structural similarities between the two phases. These Raman results led us to perform a Rietveld refinement using the structural model based on β-$K_2AgF_4$. Indeed, the results obtained from the Rietveld refinement confirmed that $Ag^I_2Ag^{II}F_4$ crystallizes in the monoclinic $P2_1/c$ space group and is isostructural to β-$K_2AgF_4$ (**Figure 1d**).

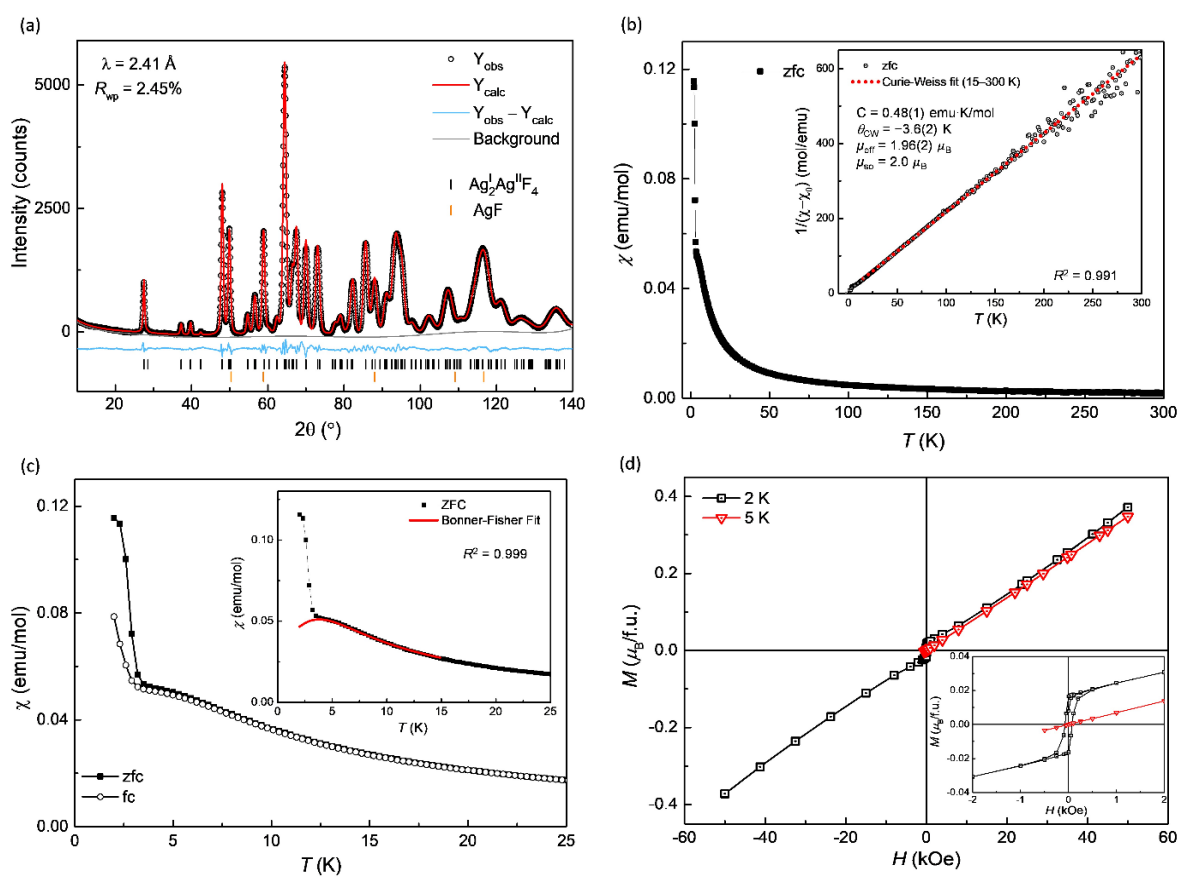

**Figure 2. (a)** The 5 K neutron powder diffraction (NPD) pattern of the ball-milled $Ag^I_2Ag^{II}F_4$ sample showing that the only impurity is small amounts of diamagnetic AgF. **(b)** Temperature-dependent magnetic susceptibility of mechanochemically prepared $Ag^I_2Ag^{II}F_4$ sample measured in a magnetic field of 1 kOe. The inset shows the Curie-Weiss fit to the inverse of magnetic susceptibility in the 15–300 K range. **(c)** Zfc and fc data in the 2–25 K temperature region showing the upturn in the magnetic susceptibility characteristic of AFM, while the Bonner-Fisher fit for the 2–15 K region is shown in the inset. **(d)** Magnetization of $Ag^I_2Ag^{II}F_4$ as a function of magnetic field measured at 2 K and 5 K, respectively. A hysteresis shown in the inset indicates a ferromagnetic ground state.

The obtained NPD data (**Figure 2a**), collected at 5 K, further confirms the structure of $Ag^I_2Ag^{II}F_4$. Traces of diamagnetic AgF impurity were also detected, in accordance with the PXRD results (**Figure 1a**). Given the structural similarities, the magnetic properties of $Ag^I_2Ag^{II}F_4$ can thus be expected to resemble those of the β-$K_2AgF_4$ phase.[43] The temperature dependent magnetic susceptibility data (**Figure 2b**) shows that $Ag^I_2Ag^{II}F_4$ behaves like a paramagnet from room temperature down to about 5 K. At about 4 K, a broad peak can be observed, indicative of low dimensional antiferromagnetic (AFM) properties after which, an abrupt increase in the magnetic susceptibility with lowering the temperature down 3 K is observed, indicative of a ferromagnetic (FM) transition.

The FM transition temperature was determined from the maximum of $-d\chi/dT$ plot to be 2.8(1) K. Both the ball milled and solid-state samples seem to show almost identical magnetic response (**Figure S6**). The inverse of the magnetic susceptibility data on the ball milled sample was corrected for a constant diamagnetic contribution and fitted from 15 K to 300 K with the Curie–Weiss law (inset in **Figure 2b**). The derived Curie constant C of 0.48(1) emu·K/mol is in good agreement with the expected value for the $S = ½$ $Ag^{2+}$ cations. A negative value for the Curie-Weiss temperature was determined, $\theta_{CW} = -3.6(2)$ K, suggesting that the sign of dominant exchange is AFM.

The 2–25 K region (**Figure 2c**) clearly shows the presence of a broad peak before the FM transition, with a maximum at about 4 K, characteristic of a 1D AFM. A fit to the region around $T_N$ (2−15 K) with a uniform AF spin-½ chain - Bonner-Fisher model[45] (inset of **Figure 1c**) is excellent and the parameters obtained, $g = 2.3$ and $|J_{1D}|/k_B = 2.9$ K are in good agreement with the values expected for an 1D AFM model, which further supports the 1D spin correlations above the $T_N$ for the $Ag^I_2Ag^{II}F_4$ phase. To get a deeper insight into the magnetism of $Ag^I_2Ag^{II}F_4$, the $M(H)$ curves was measured (**Figure 2d**). At 5 K, the magnetization displays a linear relationship with the applied magnetic field, devoid of any remnant magnetization or coercivity. However, a distinctive hysteresis loop is observed at 2 K which is below the transition temperature, characterized by a remnant magnetization of approximately 0.0165 $\mu_B$/f.u. and a coercive field of $H_c = 0.076$ kOe. Even at the largest applied magnetic field of 50 kOe, the magnetization value remains modest at 0.07 $\mu_B$/f.u. and linearly increases with the strength of the magnetic field. The magnetization value falls significantly below the anticipated saturation magnetization of $S \cdot g \cdot \mu_B \approx 1$ $\mu_B$/f.u. for the $S = ½$ spin of the $Ag^{2+}$ ion, which could be a result of a canted antiferromagnetism. A rough estimate of the canting angle (∼0.9°) in zero magnetic field could be obtained by comparing the measured remnant magnetization to the full magnetic moment of 1 $\mu_B$ per $Ag^{2+}$ ion.

Furthermore, neutron powder diffraction (NPD) data was collected on the mechanochemically synthesized $Ag^I_2Ag^{II}F_4$ sample. The measurements were performed above and below the transition temperature observed in the magnetic susceptibility. The obtained NPD data (**Figure S7**) shows only one weak feature at around 16° 2θ in the difference plot between the 1.7 and 5 K (inset in **Figure S7**). This is not surprising, as weak magnetic Bragg peaks are in fact expected due to the low magnetic dimensionality and the small moment expected from the $Ag^{2+}$ ($S = ½$). However, this sole peak was not sufficient to solve the magnetic structure of $Ag^I_2Ag^{II}F_4$.

## $Ag^IAg^{II}F_3$

The second stoichiometry that was also investigated was 1:1 molar ratio of AgF to $AgF_2$. Cooling the milling jars with liquid nitrogen was essential in stabilizing the new phase and minimizing its thermal decomposition to $Ag^I_2Ag^{II}F_4$ and $AgF_2$ - experimental details are described in the **SI**.

The laboratory PXRD data confirmed the successful formation of a new phase, presumably with the $Ag^IAg^{II}F_3$ chemical formula (**Figure 3a**). This result is in contrast to the literature reports which suggested that $Ag^IAg^{II}F_3$ should be prepared at high temperatures of around 900 °C,[26] and may thus explain the previous unsuccessful attempts.

To get a deeper insight into the reaction mechanism of the $Ag^IAg^{II}F_3$ formation, time-dependent milling experiments with sampling each 15 minutes were performed. The qualitative analysis is shown in the **SI** (**Figure S8**) while the results of the quantitative analysis are shown below (**Figure 3b**). After 15 minutes of milling, the majority of the AgF was consumed and the milled sample mainly contained of $AgF_2$, $Ag^IAg^{II}F_3$ and $Ag^I_2Ag^{II}F_4$.

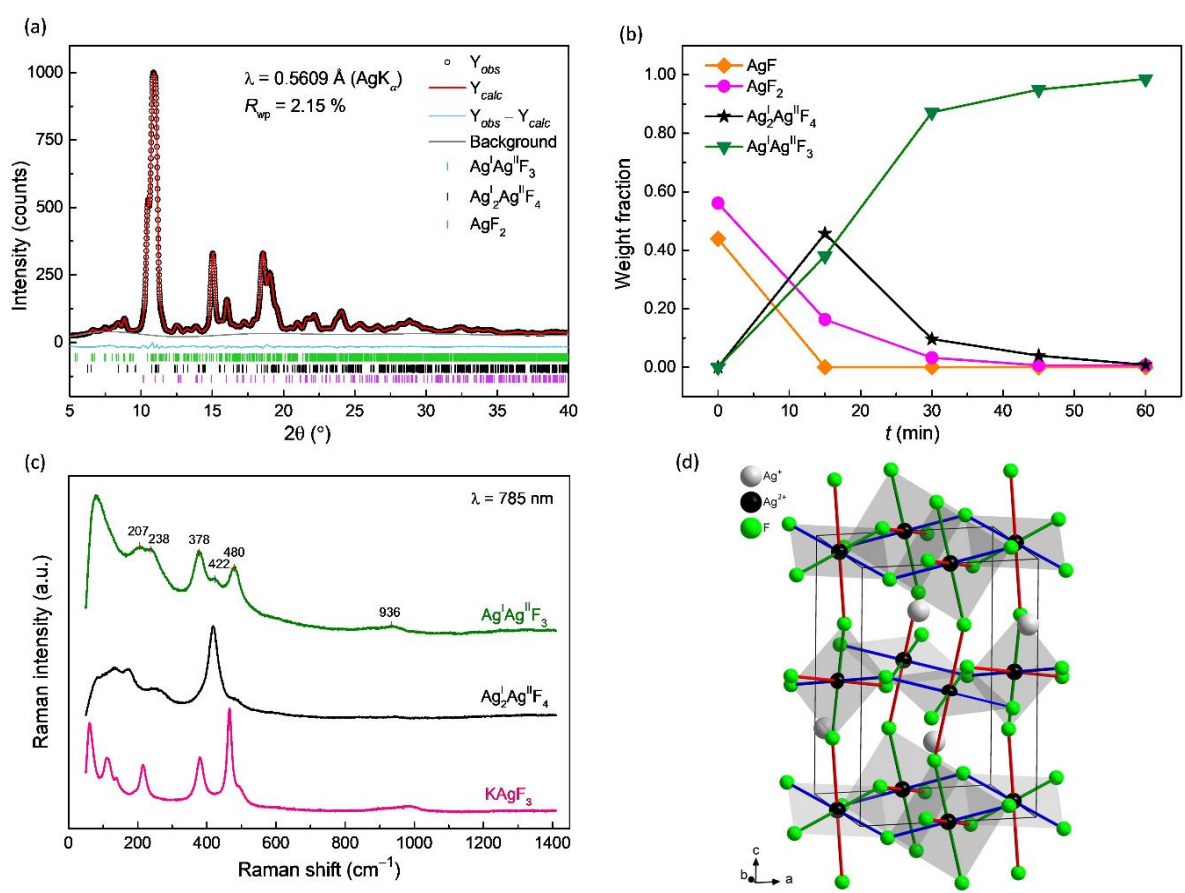

**Figure 3.** (a) Rietveld refinement of the laboratory PXRD data (Ag Kα radiation, λ = 0.5609 Å) of mechanochemically synthesized sample of $Ag^IAg^{II}F_3$ measured at room temperature; wt.% ($Ag^IAg^{II}F_3$) = 95.6(1); wt.% ($Ag^I_2Ag^{II}F_4$) = 3.3(2); wt.% ($AgF_2$) = 1.1(3). (b) Weight fraction of phases obtained from Rietveld refinement of the laboratory PXRD data as a function of milling time. The amorphous phase was not accounted in the refinement; the errors are smaller than the symbols in the graph. (c) Raman spectrum of ball-milled $Ag^IAg^{II}F_3$. For comparison, the Raman spectra of $KAgF_3$ and $Ag^I_2Ag^{II}F_4$ are also shown. (d) The crystal structure of $Ag^IAg^{II}F_3$.

Immediately after 60 minutes of milling the sample contained the $Ag^I_2Ag^{II}F_4$ phase only in trace amounts, thus it can be assumed that $Ag^IAg^{II}F_3$ is formed from the reaction of the intermediate $Ag^I_2Ag^{II}F_4$ with unreacted $AgF_2$ (**Equation 1**):

$$Ag^I_2Ag^{II}F_4 + AgF_2 \rightarrow 2Ag^IAg^{II}F_3 \qquad (1)$$

In the synchrotron PXRD data, the identified phases were $AgF_2$ and $Ag^I_2Ag^{II}F_4$, while unidentified peaks were indexed to a triclinic $P-1$ space group with a unit cell similar to $AgCuF_3$.[46] This structure was then used as a starting model for a three-phase Rietveld refinement (**Figure S9**). The crystal structure of $Ag^IAg^{II}F_3$ resembles the structures of $MAgF_3$ compounds ($M$ = K, Rb and Cs)[47] with additional distortions which lower the symmetry. Consequently, similarities were expected between the Raman spectra of these phases. The Raman spectra of $Ag^IAg^{II}F_3$ (**Figure 3c**) was therefore assigned using the literature reports available on $KAgF_3$.[48] For comparison, a solid-state method was also applied to obtain the $KAgF_3$ phase (see **Experimental** section, **SI** for details) and its Raman spectrum compared to the Raman spectrum of $Ag^IAg^{II}F_3$. The peak at 480 cm$^{-1}$ could be assigned to the $[AgF_6]^{4-}$ octahedron stretching band ($B_{2g}$), peak at 378 cm$^{-1}$ to the $[AgF_2]$ plane stretching ($A_{1g}$) and the peak at 207 cm$^{-1}$ to the $[AgF_6]^{4-}$ octahedron bending ($E_g$). Two broad, less intense peaks located between 850–950 cm$^{-1}$ are also observed in the literature throughout the $MAgF_3$ series, but were either assigned to an impurity, or to a combination of two Raman active modes. The peaks at 238 cm$^{-1}$ and 936 cm$^{-1}$ could not be assigned to any vibrational mode. It's noteworthy that the most intense Raman active mode of $Ag^I_2Ag^{II}F_4$ at 422 cm$^{-1}$ is also present in the Raman spectrum of $Ag^IAg^{II}F_3$, albeit with a weak signal due to its low content - 3 wt.% as determined by Rietveld refinement (**Figure 3a**). The presence of the 2:1 phase could be explained through the observation that the mechanochemically synthesized $Ag^IAg^{II}F_3$ sample, stored in the glovebox at the room temperature, decomposes over time into $AgF_2$ and $Ag^I_2Ag^{II}F_4$ (**Figure S10**).

A Rietveld refinement analysis of the PXRD data was successfully performed only by using the triclinic $P-1$ space group, which confirmed that $Ag^IAg^{II}F_3$ is isostructural to $AgCuF_3$ and $NaCuF_3$.[46] In this structure, there are four non-equivalent silver(II) sites, all coordinated with six fluorine atoms, forming a distorted $[AgF_6]^{4-}$ octahedron. Each octahedron has two short, two middle and two long Ag−F contacts. Short and middle bond lengths range from 2.072(16) Å to 2.161(13) Å and longer bond lengths from 2.403(11) Å to 2.441(16) Å (**Table S3**). Since the two short and the two middle Ag−F bonds have similar lengths, the octahedra exhibits axial elongation (4 + 2) and therefore could be Jahn-Teller active (**Figure 3d**). The corner-sharing $[AgF_6]^{4-}$ units form infinite chains along the $c$-axis, while in the $ab$ plane they form puckered $[AgF_2]$ sheets, resembling the structural arrangement of the $MAgF_3$ compounds ($M$ = K, Rb and Cs)[47] (**Figure S11**). In the crystal structure of $Ag^IAg^{II}F_3$, there are two non-equivalent silver(I) atoms, both coordinated with seven fluorine atoms. Together they form a double fluorine-bridged $[Ag_2F_{12}]^{10-}$ dimers (**Figure S12**) with the length of the Ag−F bond ranging from 2.341(16) Å to 3.072(20) Å. The BVS analysis confirms the valence state of silver(I) and silver(II) atoms (**Table S4**), and offers further support for the correctness of the crystal structures.

The magnetic susceptibility data (**Figure 4a**) reveals intriguing features that provide valuable insights into the magnetic properties of $Ag^IAg^{II}F_3$. An $AgF_2$ impurity signal at approximately 164 K is observed and quantified to about 0.8(3) wt% which is in good agreement with the value obtained of 1.1(2) wt.% from the Rietveld refinement - **Figure 3a**. The presence of $AgF_2$ is expected due to the 10 mol% excess of $AgF_2$ that was added for the synthesis. A weak FM signal is also observed which corresponds to the 2:1 phase. It should be noted that if the magnetization was measured on a freshly prepared $Ag^IAg^{II}F_3$ sample from a stoichiometric 1:1 ratio $AgF: AgF_2$ a smaller amount of $AgF_2$ is present but a larger amount of the 2:1 phase is observed as a secondary phase which is minimized when an excess of $AgF_2$ is used during the synthesis.

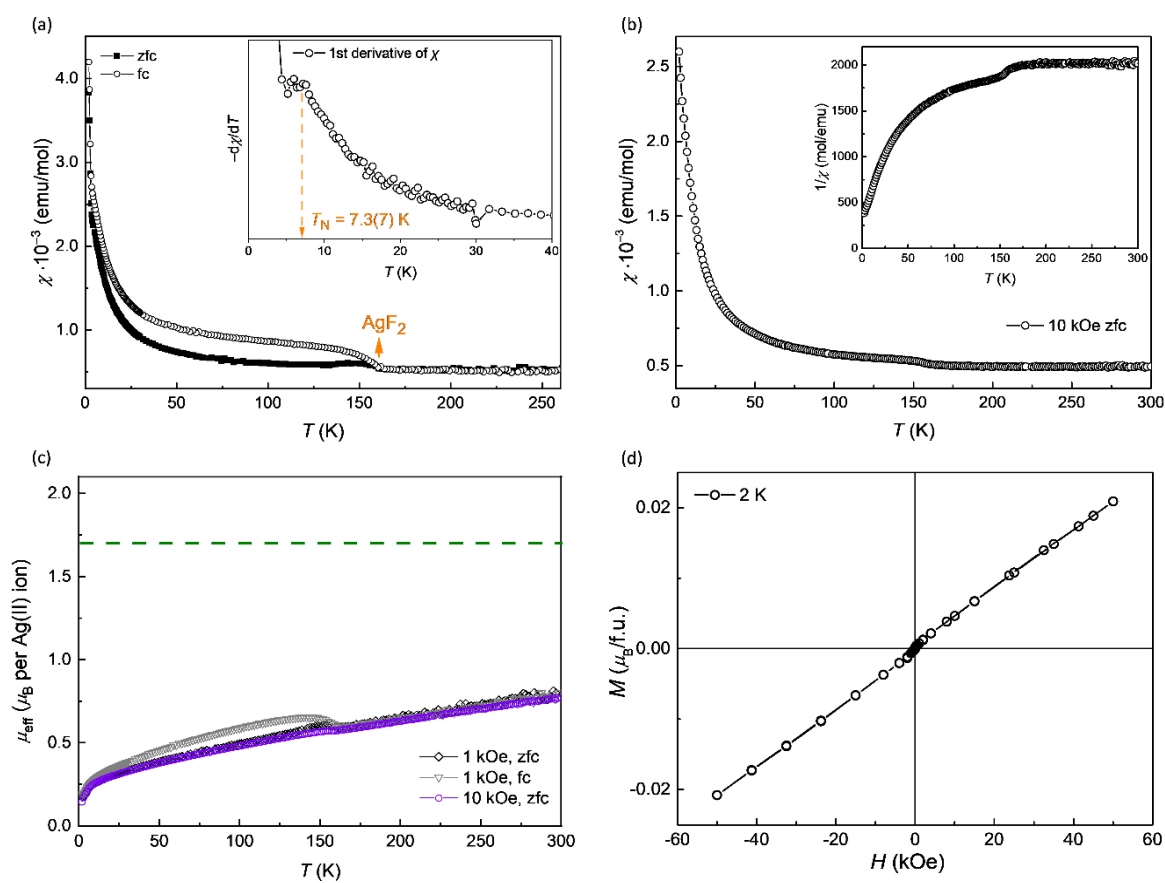

**Figure 4.** Temperature-dependent susceptibility, χ, of $Ag^IAg^{II}F_3$, measured in a magnetic field of 1 kOe **(a)** and 10 kOe **(b).** The inset in (**a**) shows the first derivative indicating a deviation from the Currie-Weiss law at 7.3(7) K. The inset in (**b**) shows the inverse of the magnetic susceptibility (fc) measured at 10 kOe. **(c)** Plot of the effective magnetic moment, $\mu_{eff}$, as a function of temperature for field cooled and zero field cooled curves at 1 kOe and for the field-cooled curve measured at 10 kOe. **(d)** Magnetization of $Ag^IAg^{II}F_3$ as a function of magnetic field, measured at 2 K.

The first derivative of magnetic susceptibility indicates the temperature where the deviation from the Curie–Weiss law is observed. The magnetic susceptibility measurement was also repeated in ten times larger magnetic field of 10 kOe to minimize the influence of the ferromagnetic signal of $AgF_2$. Further analysis of the magnetic susceptibility measured at 10 kOe (**Figure 4b**) in the high-temperature range between 200–300 K uncovers a persistent decrease of the susceptibility with decreasing temperature, indicative of an antiferromagnetic interactions persisting even at elevated temperatures. The observed behavior bears resemblance to $KAgF_3$, a one-dimensional antiferromagnet.[49] This hypothesis is also supported by the fact that effective magnetic moment at 300 K is well below the theoretical expected value of 1.7 $\mu_B$ since it reaches only 1 $\mu_B$ (**Figure 4c**).

The effective magnetic moment also decreases with decreasing temperature and reaches value of approximately 0.15 $\mu_B$, indicating that antiferromagnetic interactions are predominant throughout the whole measured temperature range. Furthermore, the magnetization measurements (**Figure 4d**) revealed that the magnetic moments do not attain saturation even under the influence of high magnetic fields, a characteristic that strongly suggests the presence of antiferromagnetic interactions.

**CONCLUSIONS**

In this study, mechanochemistry was introduced as a new synthetic method for fluoridoargentate(II) and provided the experimental evidence for the first binary mixed-valent silver(I,II) fluorides, $Ag^I_2Ag^{II}F_4$ and $Ag^IAg^{II}F_3$. While the $Ag^I_2Ag^{II}F_4$ phase is stable at room temperature, the $Ag^IAg^{II}F_3$ seems to be metastable and required milling at liquid nitrogen temperatures. Synchrotron PXRD and Raman spectroscopy studies confirmed the post-perovskite structure of $Ag^I_2Ag^{II}F_4$, which resembles that of β-$K_2AgF_4$. This compound shows a canted 1D antiferromagnetism below 2.8(1) K. On the other side, the $Ag^IAg^{II}F_3$ phase seems to be metastable and shares structural features with $AgCuF_3$ and the $M^IAgF_3$ series showing characteristics of a one-dimensional antiferromagnet.

This work demonstrates the effectiveness of the mechanochemical approach in the synthesis of fluoridoargentates(II), offering the possibility for a low-temperature synthesis which overcomes the challenges of $Ag^{2+}$ reactivity and thermal instability of $AgF_2$. This will certainly pave the way for the discovery of new silver(II) compounds.

**ACKNOWLEDGMENTS**

Support from the Slovenian Research and Innovation Agency (J2-2496, P2-0105, P2-0348) and the European Research Council Starting Grant (950625) under the European Union's Horizon 2020 Research and Innovation Programme; and the Jožef Stefan Institute Director's Fund, are gratefully acknowledged. This project has received funding from the European Union's Horizon 2020 Research and Innovation Programme under the Marie Skłodowska-Curie grant agreement No. 101031415.

Part of the research described in this paper was performed at the Canadian Light Source, a national research facility of the University of Saskatchewan, which is supported by the Canada Foundation for

**SUPPORTING INFORMATION**

The Experimental procedures, description of the characterization techniques, together with selected results are included in the **Supporting Information File**. Deposition numbers 2321112 (for $Ag_2F_3$) and 2321184 (for $Ag_3F_4$) contain the supplementary crystallographic data for this paper. These data are provided free of charge by the joint Cambridge Crystallographic Data Centre and Fachinformationszentrum Karlsruhe Access Structures service.

# Supporting Information

## Mechanochemical Synthesis and Magnetic Properties of the Mixed-Valent Binary Silver(I,II) Fluorides, $Ag^I_2Ag^{II}F_4$ and $Ag^IAg^{II}F_3$

Matic Belak Vivod[a,b], Zvonko Jagličić[c,d], Graham King[e], Thomas C. Hansen[f], Matic Lozinšek[a,b], and Mirela Dragomir[a,b]*

[a] *Jožef Stefan Institute, Jamova cesta 39, 1000 Ljubljana, Slovenia*
[b] *Jožef Stefan International Postgraduate School, Jamova cesta 39, 1000 Ljubljana, Slovenia.*
[c] *Institute of Mathematics, Physics and Mechanics, 1000 Ljubljana, Slovenia*
[d] *Faculty of Civil and Geodetic Engineering, University of Ljubljana, Jamova cesta 2, 1000 Ljubljana, Slovenia*
[e] *Canadian Light Source, 44 Innovation Blvd, Saskatoon, SK S7N 2V3, Canada*
[f] *Institut Laue-Langevin, 38042 Grenoble Cedex 9, France*

*Corresponding author: mirela.dragomir@ijs.si

SECTION A     Synthetic Procedures
SECTION B     Characterization
SECTION C     Results
SECTION D     References

# SECTION A    Synthetic Procedures

**Materials**

$Ag^I_2Ag^{II}F_4$, $Ag^IAg^{II}F_3$, $K_2AgF_4$ and $KAgF_3$ were synthesized from commercially available $AgF_2$ (98% Sigma-Aldrich or 99% Fluorochem), AgF (99% Thermo Scientific or 99.9% Sigma-Aldrich) and KF (99% Fluorochem), which were used as received. Handling of all chemicals was performed in an argon-filled glove box (M. Braun, Germany) with water and oxygen levels below 0.1 ppm.

The obtained polycrystalline samples were also handled in a glove box under inert conditions at all times.

**Mechanochemical synthesis**

For the mechanochemical synthesis of $Ag^I_2Ag^{II}F_4$, approximately 1 g of a 2:1 molar mixture of AgF and $AgF_2$ was homogenised in an agate mortar for 15 minutes and placed together with a tungsten carbide milling ball ($d = 15$ mm) into a 10 ml tungsten carbide milling jar equipped with a PTFE gasket. The BRR (ball-to-reagent ratio) was set to 5. Milling was performed in a mixer mill Retsch, MM 400. The powder mixture was milled for a total of 60 min (2 cycles), each cycle consisting of 15 minutes milling followed by 15 minutes of cooling. The same reaction conditions were used for the synthesis of β-$K_2AgF_4$ but with a 10 mol% excess of $AgF_2$ and a total milling time of 90 minutes (3 cycles).

For the mechanochemical synthesis of $Ag^IAg^{II}F_3$, 1 g of a homogenised powder of 1:1 or 1:1.1 molar ratio AgF:$AgF_2$ was placed together with a stainless-steel milling ball ($d = 15$ mm, BRR = 12) into a 10 ml stainless steel milling jar equipped with a PTFE gasket. The milling was carried out in a Retsch MM 400. The powder mixture was milled for 60 min - 3 cycles, each consisting of 15 minutes milling and approximatively 5 minutes of cooling by immersing the jars in liquid nitrogen until equilibrium.

**Solid-state synthesis**

In a typical solid-state synthesis of $Ag^I_2Ag^{II}F_4$ and $Ag^IAg^{II}F_3$, a stoichiometric mixture of AgF and $AgF_2$ (approx. 300 mg) was hand homogenised in agate mortar and then placed in a platinum crucible, which was enclosed in a nickel reactor and heated for 3 days at 300 °C, followed by natural cooling to room temperature. When the reactions were performed at higher temperatures, sample decomposition was observed. For the comparison of Raman spectra, $KAgF_3$ was prepared according to the literature report.[50]

# SECTION B    Characterization

**Powder X-ray diffraction**

The homogenized samples were transferred to dried quartz capillaries ($d = 0.3$ mm, Hilgenberg) that were then flame sealed.

Laboratory powder X-ray diffraction patterns were measured with a Rigaku OD XtaLAB Synergy-S, Dualflex diffractometer equipped with Eiger2 R CdTe 1M hybrid pixel array detector using Ag Kα radiation ($\lambda = 0.56087$ Å) at 65 kV, 0.67 mA. The data were collected at room temperature or 100 K, in transmission mode, with detector distance of 90 mm and the phases were identified using X-Pert High Score Plus[51] and PDF-4 + database. The powder patterns were processed and extracted by CrysAlisPro.[52] The synchrotron X-ray diffraction measurements were measured in quartz capillaries ($d = 0.3$ mm, Hilgenberg) at the Canadian Light Source (CLS) at the Brockhouse High Energy Wiggler Beamline using $\lambda = 0.3497$ Å radiation and a Varex area detector. The indexing, structural refinements and quantitative phase composition analyses were all performed using GSAS-II software.[53]

**Neutron powder diffraction**

Approximately 1 g of the mechanochemically synthesised $Ag^I_2Ag^{II}F_4$ was loaded into a vanadium can inside the glovebox. The sample was measured at the high intensity powder diffractometer D20 at the Institute Laue-Langevin (ILL) with $\lambda = 2.41$ Å wavelength.[54]

**Raman spectroscopy**

Raman spectra were measured in quartz capillaries at room-temperature on a Bruker Senterra II confocal Raman microscope using emission line (785 nm) with the power output of 1 mW. The spectra were measured in the range from 50 to 1410 cm$^{-1}$ with a resolution of 1.5 cm$^{-1}$. The green emission line (532 nm, 12.5 mW) was used for the study of photochemical degradation of $AgF_2$.

**Magnetisation**

Magnetic susceptibility measurements were performed with a Quantum Design MPMS-XL-5 SQUID magnetometer. Susceptibility as a function of temperature $T$ was measured between 2–300 K in a constant magnetic field of 1 kOe. The isothermal magnetization was measured between −50 and +50 kOe at temperature of 2 K. The data were corrected for the experimentally determined contribution of quartz capillary and the diamagnetic response of the compound due to closed atomic shells as obtained from Pascal's tables.[55]

**SECTION C    Results**

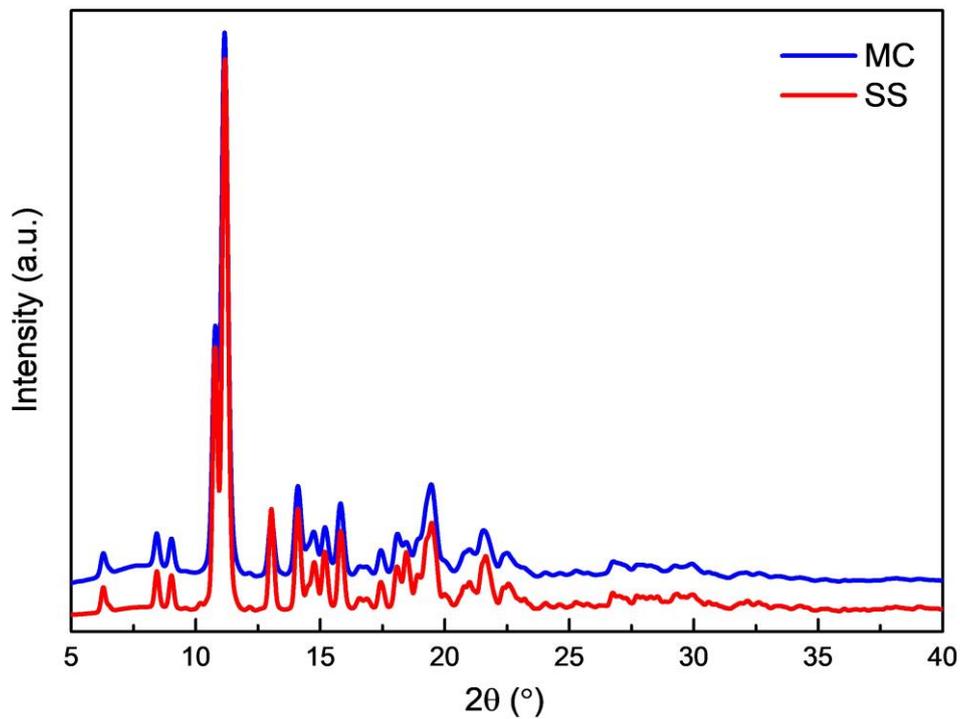

**Figure S1.** Laboratory PXRD data of Ag$^I_2$Ag$^{II}$F$_4$ prepared by mechanochemistry (MC) and solid-state synthesis (SS), measured with Ag Kα radiation at room temperature.

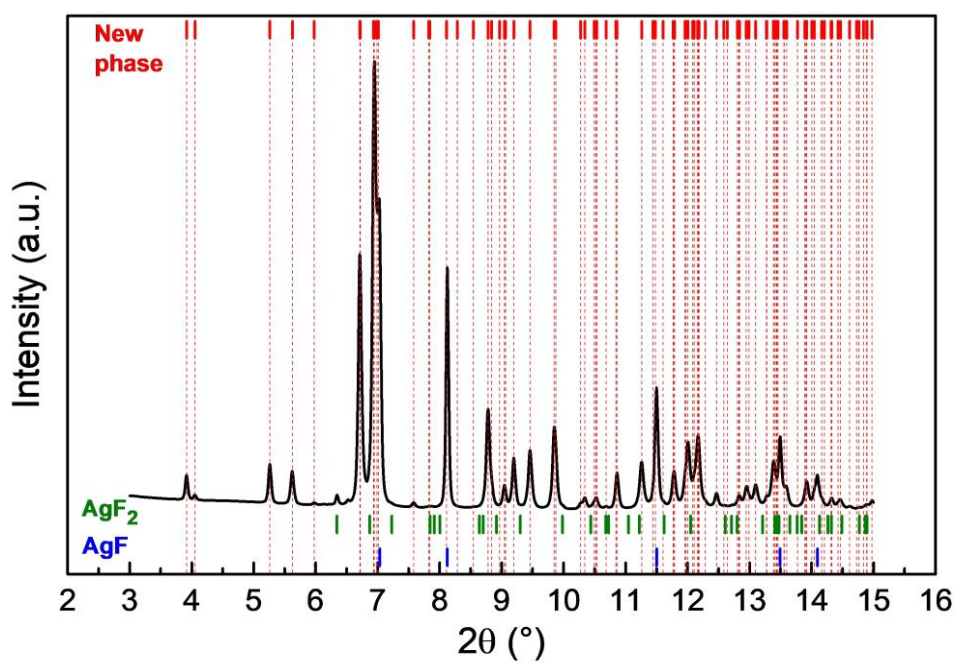

**Figure S2**. Unit cell indexing ($P2_1/c$ space group) of the synchrotron PXRD data measured with λ = 0.3497 Å.

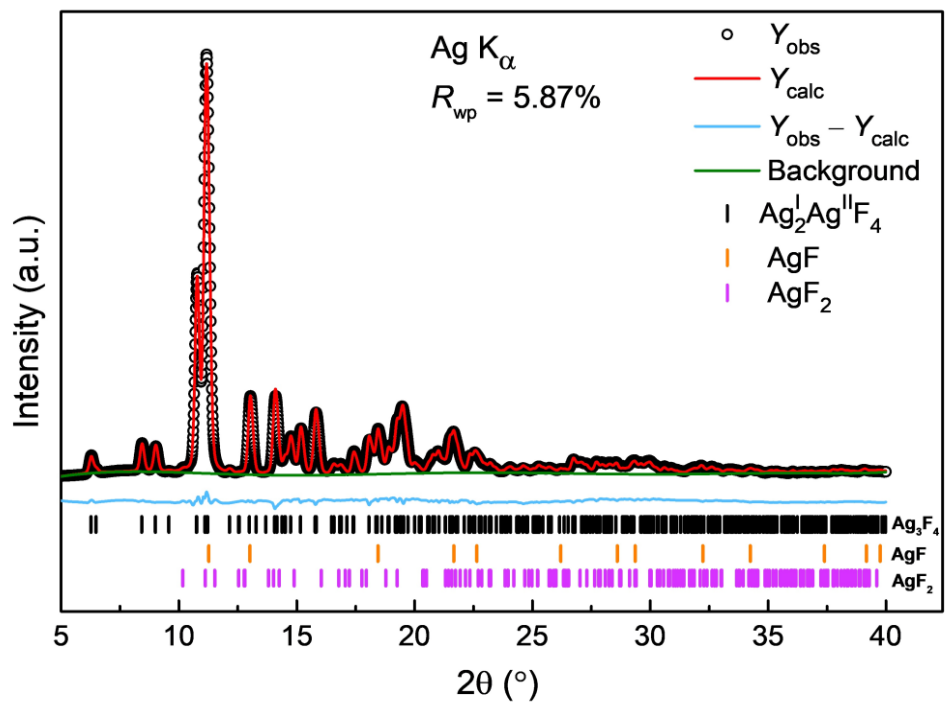

**Figure S3.** Rietveld refinement of the laboratory PXRD data (Ag Kα radiation) of solid-state sample of $Ag^I_2Ag^{II}F_4$ measured at room temperature. Wt.% ($Ag^I_2Ag^{II}F_4$) = 85.4(2); wt.% (AgF) = 14.3(1); wt.% ($AgF_2$) = 0.3(4).

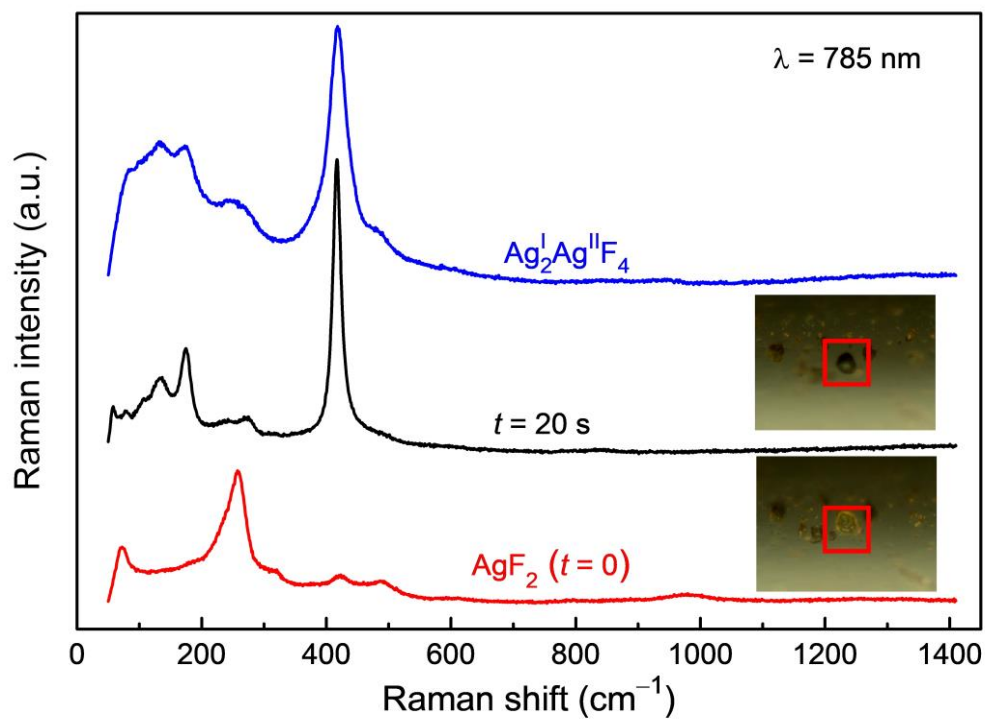

**Figure S4.** Raman spectra measured using 785 nm excitation laser with 1 mW power. Bottom Raman spectrum (red) corresponds to $AgF_2$, middle spectrum (black) was obtained after $AgF_2$ was illuminated with green laser (532 nm, 12.5 mW) for 20 seconds, and top spectrum (blue) was obtained on the mechanochemically synthesised $Ag^I_2Ag^{II}F_4$.

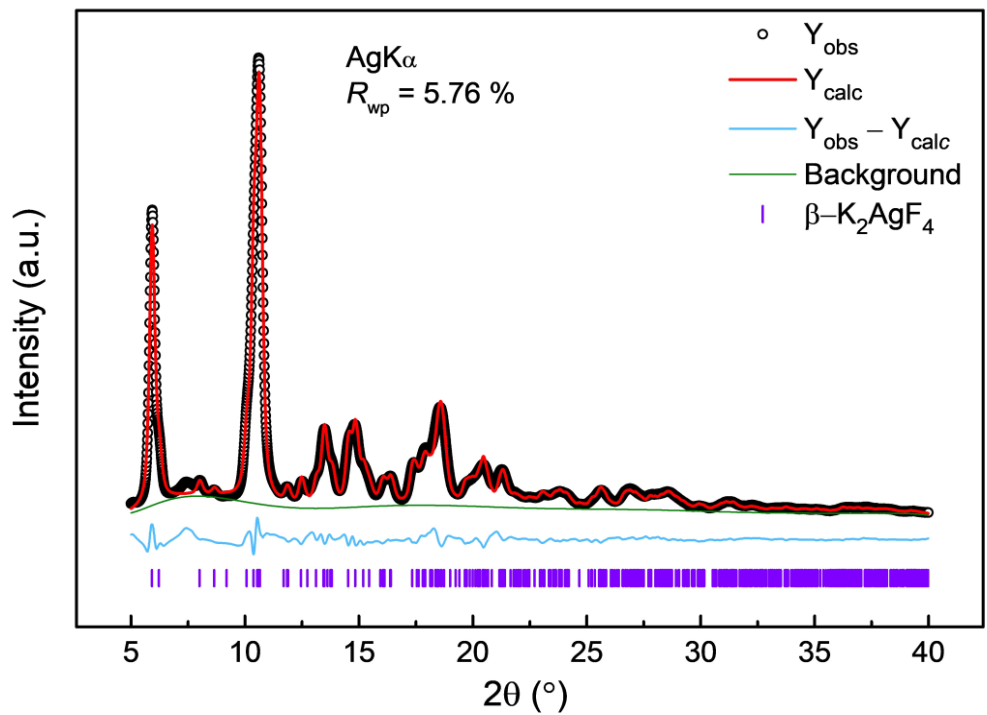

**Figure S5.** Single-phase Rietveld refinement of the laboratory PXRD data (Ag Kα radiation, λ = 0.5609 Å) of mechanochemically synthesised sample of β-$K_2AgF_4$ measured at room temperature. Note that there is one unidentified peak at around 7.5° 2θ.

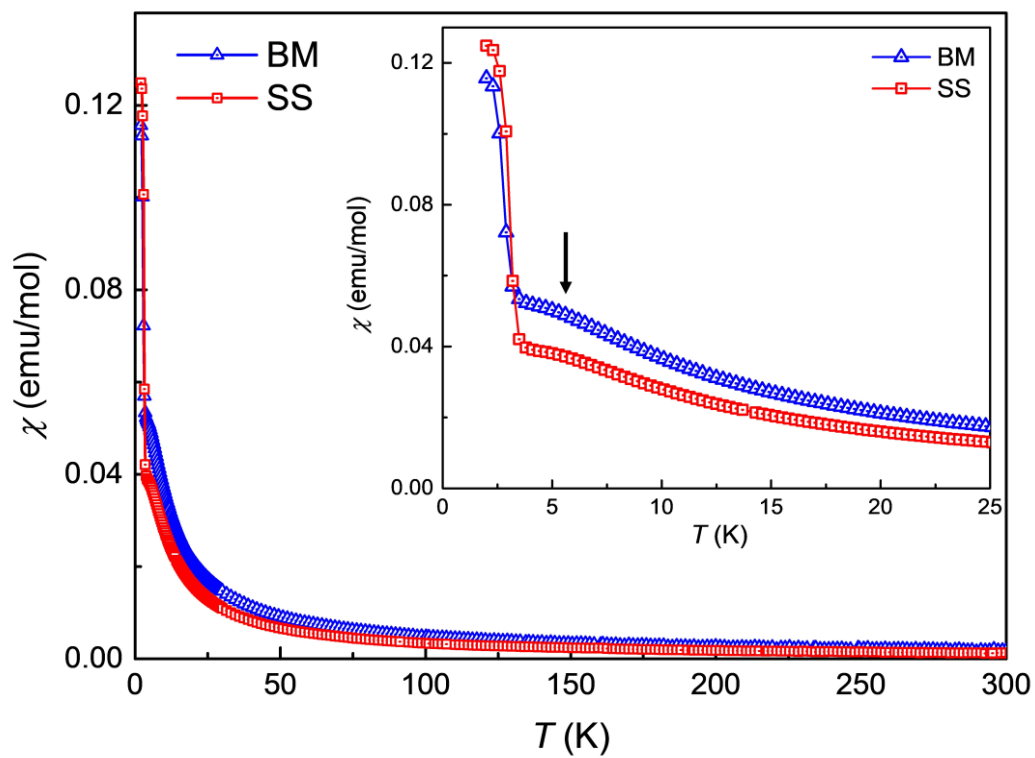

**Figure S6.** Zero field-cooled curves of temperature-dependent magnetic susceptibility of $Ag^I_2Ag^{II}F_4$ prepared with ball milling (BM) and solid-state synthesis (SS).

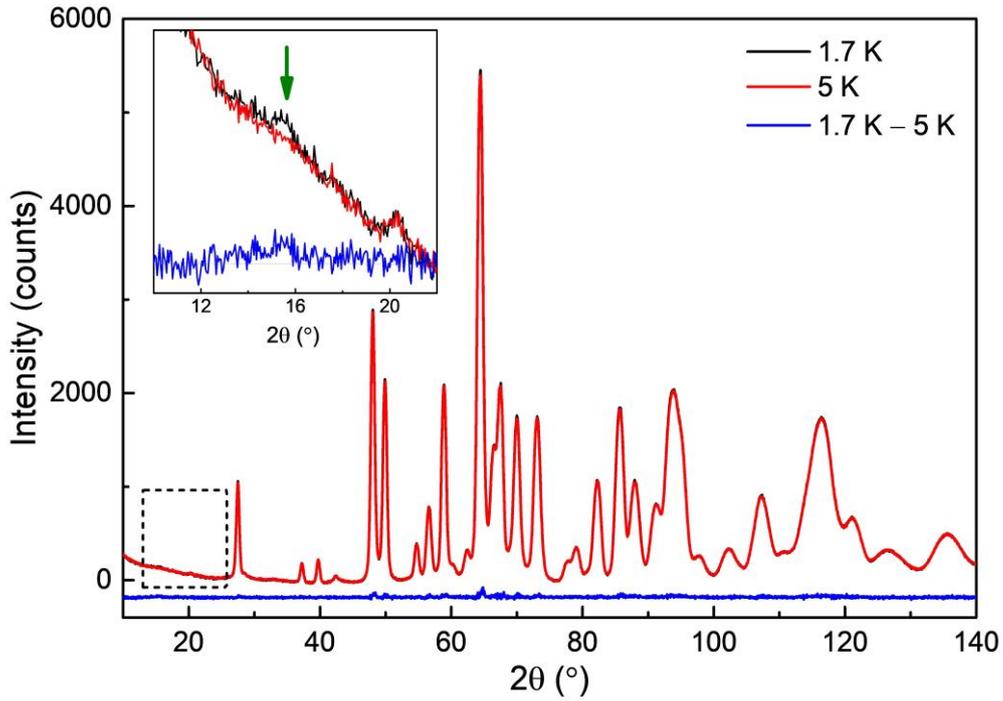

**Figure S7**. The NPD data measured at 5 and 1.7 K and the difference plot. A magnetic peak at about 16° 2θ can be observed (green arrow in the inset).

**Table S1**: Crystallographic parameters of $Ag^I_2Ag^{II}F_4$ obtained by the Rietveld refinement of synchrotron data.

| Space group | $P2_1/c$ (no. 14) |
|---|---|
| Chemical formula weight (g/mol) | 399.6 |
| Z | 2 |
| *a*, *b*, *c* (Å) | 3.5691(1), 9.8866(2), 5.9829(2) |
| β (°) | 92.83(1) |
| V (Å$^3$) | 210.86(1) |
| Ag1 (*x*, *y*, *z*) | 0, 0, 0 |
| $U_{iso}$ | 0.0110(4) |
| Oxidation state | +2 |
| Ag2 (*x*, *y*, *z*) | 0.5122(3), 0.1786(2), 0.4304(2) |
| $U_{iso}$ | 0.0162(3) |
| Oxidation state | +1 |
| F1 (*x*, *y*, *z*) | 0.557(2), 0.452(1), 0.267(1) |
| $U_{iso}$ | 0.053(3) |
| F2 (*x*, *y*, *z*) | 0.016(2), 0.197(1), 0.139(1) |
| $U_{iso}$ | 0.021(2) |
| GOF | 1.62 |
| $R_{wp}$ | 0.0255 |

**Table S2.** Interatomic distances and bond valence sum analysis[56] for silver atoms in $Ag^I_2Ag^{II}F_4$. The following bond valence parameters were employed: $b = 0.37$ Å; $R_0 = 1.80$ Å ($Ag^I$–F) and 1.79 Å ($Ag^{II}$–F).[57,58]

| Interatomic distances (Å) | |
|---|---|
| Ag1−F1 × 2 | 2.111(6) |
| Ag1−F1 × 2 | 2.119(8) |
| Ag1−F2 × 2 | 2.529(7) |
| **Bond valence sum** | |
| Ag1 | 1.94 |
| **Interatomic distances (Å)** | |
| Ag2−F1 | 2.391(7) |
| Ag2−F1 | 2.431(6) |
| Ag2−F1 | 2.465(7) |
| Ag2−F2 | 2.536(7) |
| Ag2−F2 | 2.537(8) |
| Ag2−F2 | 2.571(7) |
| Ag2−F2 | 2.884(8) |
| **Bond valence sum** | |
| Ag2 | 1.00 |

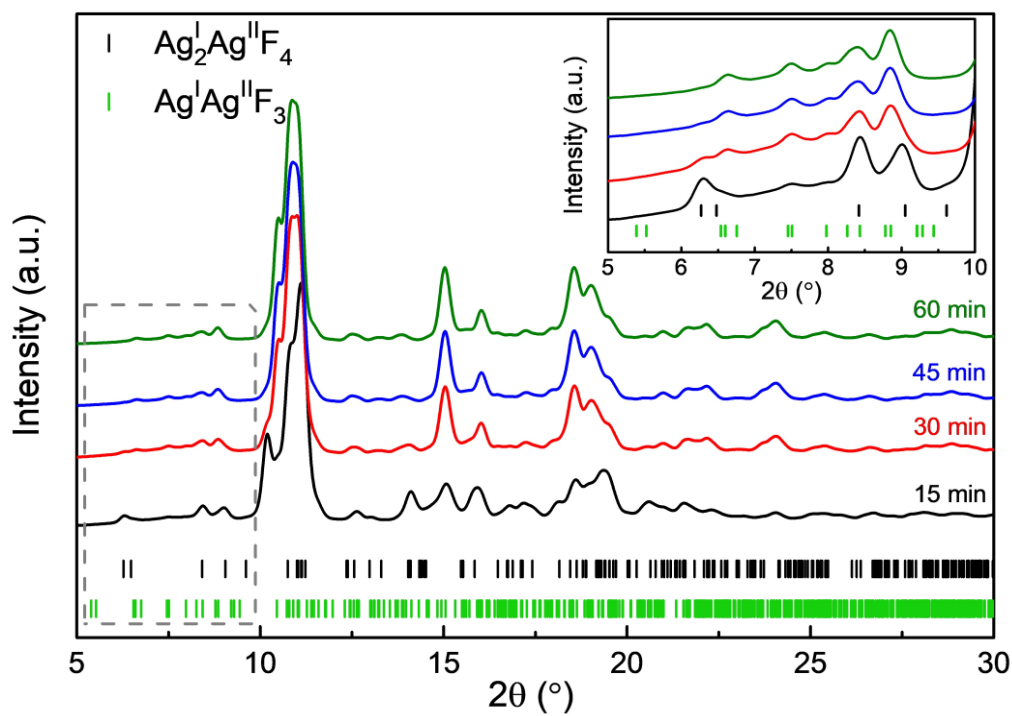

**Figure S8.** Time-dependent sampling during the mechanochemical synthesis of $Ag^IAg^{II}F_3$. It can be seen that the $Ag^I_2Ag^{II}F_4$ possibly forms concomitantly with the $Ag^IAg^{II}F_3$ phase and gets consumed as the reaction proceeds.

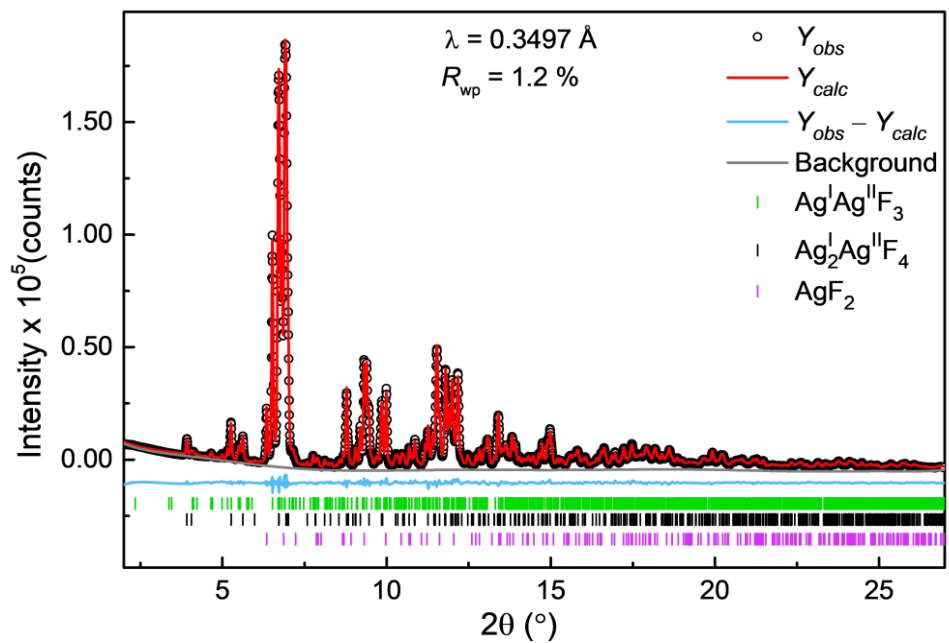

**Figure S9.** Three-phase Rietveld refinement of a synchrotron data collected at room temperature on the $Ag^IAg^{II}F_3$ sample prepared by solid-state method; wt.% ($Ag^IAg^{II}F_3$) = 57.4(1) %; wt.% ($Ag^I_2Ag^{II}F_4$) = 39.3(1) %; wt.% ($AgF_2$) = 3.3(1) %.

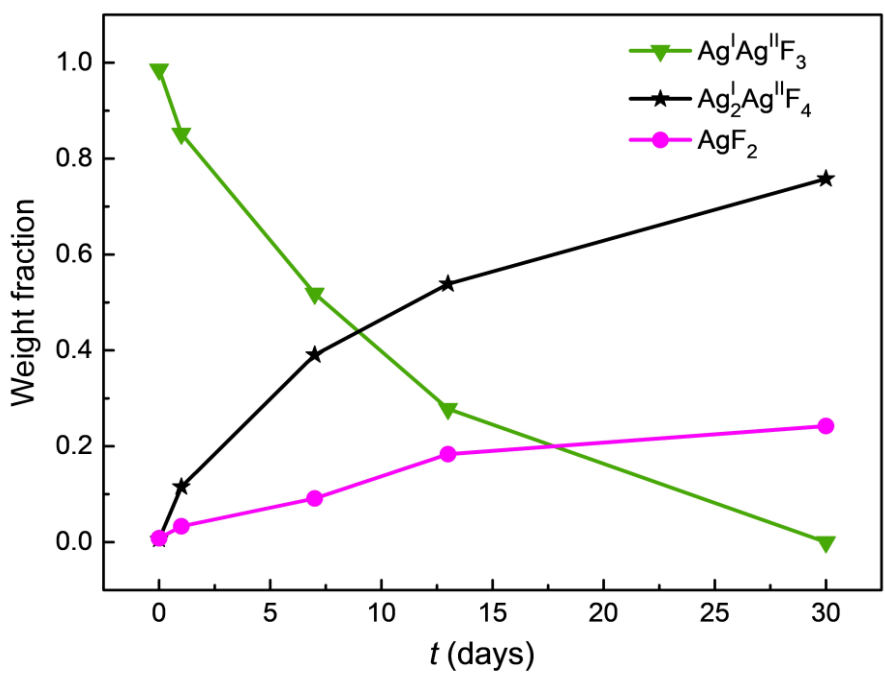

**Figure S10.** Evolution of the phase composition decomposition of mechanochemically synthesized $Ag^IAg^{II}F_3$ stored in the glove box as a function of time over a 30-day period. Weight fraction of phases were obtained by Rietveld refinement of the laboratory PXRD data measured at room temperature. The amorphous phase was not accounted in the refinement; therefore, the amount of crystalline phase is overestimated. The errors bars are smaller than the symbols in the graph.

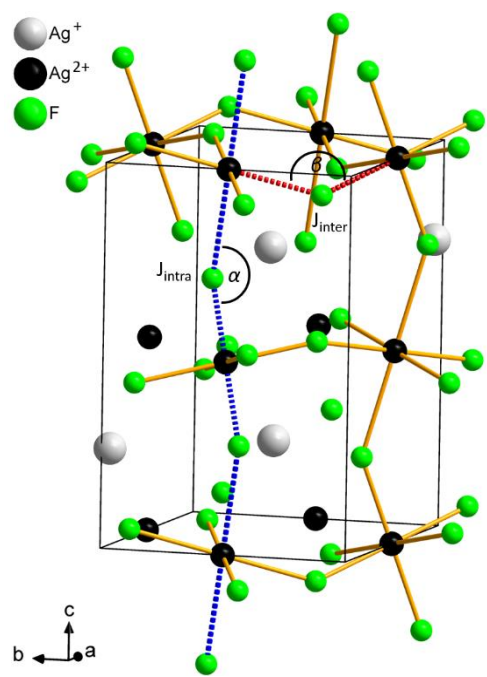

**Figure S11**. Unit cell of $Ag^IAg^{II}F_3$ showing the corner-sharing $[AgF_6]^{4-}$ units that form infinite chains along the *c*. The intrachain (blue) and interchain (red) magnetic coupling are also displayed. The angle α is 149.5(7)° while β is 138.4(6)°.

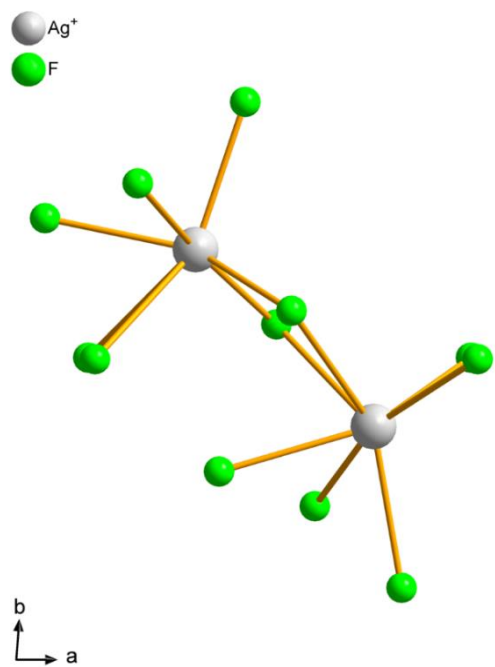

**Figure S12**. An $[Ag_2F_{12}]^{10-}$ dimer in $Ag^IAg^{II}F_3$.

**Table S3**. Crystallographic parameters of Ag$^{I}$Ag$^{II}$F$_3$ obtained by the Rietveld refinement of synchrotron data.

| | Ag$^{I}$Ag$^{II}$F$_3$ |
|---|---|
| **Space group** | $P{-}1$ (no. 2) |
| **Chemical formula weight (g/mol)** | 272.73 |
| **Z** | 1 |
| **a, b, c (Å)** | 5.9577(1), 5.8217(1), 8.5467(2) |
| **α, β, γ (°)** | 91.565(2), 90.644(2), 85.977(1) |
| **V (Å$^3$)** | 295.58(1) |
| **Ag1 (x, y, z)** | 0.0515(8), 0.9834(8), 0.2550(5) |
| $U_{iso}$ | 0.020(2) |
| **Oxidation state** | +1 |
| **Ag2 (x, y, z)** | 0.5501(8), 0.5066(8), 0.2444(5) |
| $U_{iso}$ | 0.025(2) |
| **Oxidation state** | +1 |
| **Ag3 (x, y, z)** | 0.5, 0, 0 |
| $U_{iso}$ | 0.006(3) |
| **Oxidation state** | +2 |
| **Ag4 (x, y, z)** | 0, 0.5, 0 |
| $U_{iso}$ | 0.005(3) |
| **Oxidation state** | +2 |
| **Ag5 (x, y, z)** | 0.5, 0, 0.5 |
| $U_{iso}$ | 0.009(3) |
| **Oxidation state** | +2 |
| **Ag6 (x, y, z)** | 0, 0.5, 0.5 |
| $U_{iso}$ | 0.009(3) |
| **Oxidation state** | +2 |
| **F7 (x, y, z)** | 0.064(3), 0.594(3), 0.733(1) |
| $U_{iso}$ | 0.021(6) |
| **F8 (x, y, z)** | 0.212(3), 0.218(3), 0.559(3) |
| $U_{iso}$ | 0.042(7) |
| **F9 (x, y, z)** | 0.204(2), 0.198(2), -0.078(2) |
| $U_{iso}$ | 0.008(6) |
| **F10 (x, y, z)** | 0.310(3), 0.729(3), 0.430(2) |
| $U_{iso}$ | 0.014(6) |
| **F11 (x, y, z)** | 0.291(2), 0.670(3), 0.046(3) |
| $U_{iso}$ | 0.016(7) |
| **F12 (x, y, z)** | 0.562(3), 0.893(3), 0.771(1) |
| $U_{iso}$ | 0.014(6) |
| **GOF** | 2.16 |

**Table S4.** Interatomic distances and bond valence sum analysis[56] for silver atoms in $Ag^{I}Ag^{II}F_3$. The following bond valence parameters were employed: $b = 0.37$ Å; $R_0 = 1.80$ Å ($Ag^{I}$–F) and 1.79 Å ($Ag^{II}$–F).[57,58]

| Interatomic distances (Å) | |
|---|---|
| Ag1–F9 | 2.409(15) |
| Ag1–F12 | 2.473(17) |
| Ag1–F7 | 2.508(17) |
| Ag1–F10 | 2.560(17) |
| Ag1–F8 | 2.607(19) |
| Ag1–F11 | 2.843(18) |
| Ag1–F8 | 3.072(20) |
| Ag2–F7 | 2.341(16) |
| Ag2–F11 | 2.430(17) |
| Ag2–F10 | 2.451(19) |
| Ag2–F12 | 2.464(18) |
| Ag2–F9 | 2.745(19) |
| Ag2–F8 | 2.772(15) |
| Ag2–F11 | 2.806(20) |
| **Bond valence sum** | |
| Ag1 | 0.84 |
| Ag2 | 0.97 |

| Interatomic distances (Å) | |
|---|---|
| Ag3–F12 × 2 | 2.066(11) |
| Ag3–F9 × 2 | 2.147(13) |
| Ag3–F11 × 2 | 2.407(16) |
| Ag4–F11 × 2 | 2.081(15) |
| Ag4–F9 × 2 | 2.161(13) |
| Ag4–F7 × 2 | 2.403(11) |
| Ag5–F10 × 2 | 2.078(16) |
| Ag5–F8 × 2 | 2.115(16) |
| Ag5–F12 × 2 | 2.435(11) |
| Ag6–F7 × 2 | 2.072(16) |
| Ag6–F8 × 2 | 2.087(11) |
| Ag6–F10 × 2 | 2.441(16) |
| **Bond valence sum** | |
| Ag3 | 2.09 |
| Ag4 | 2.03 |
| Ag5 | 2.10 |
| Ag6 | 2.18 |

# SECTION D     References